\documentclass{kapproc} 

\usepackage{procps}

\usepackage[dvips]{graphicx}

\upperandlowercase

\kluwerbib 

\begin{document}

\articletitle[Dust Tori in AGN]
{Penetrating Dust Tori in AGN}

\author{Gabriela Canalizo\altaffilmark{1}, Claire Max\altaffilmark{2,3}, 
Robert Antonucci\altaffilmark{4}, David Whysong\altaffilmark{4},
Alan Stockton\altaffilmark{5}, Mark Lacy\altaffilmark{6}} 
 
\affil{\altaffilmark{1} Department of Earth Sciences and IGPP, 
University of California, Riverside, CA 95521, 
\altaffilmark{2}IGPP, Lawrence 
Livermore National Laboratory,
\altaffilmark{3}Center for Adaptive Optics, University of California, 
Santa Cruz, 
\altaffilmark{4}Physics Department, University of California, Santa Barbara, 
\altaffilmark{5}Institute for Astronomy, University of Hawaii, 
\altaffilmark{6}Spitzer Science Center, MS 220-6, Caltech}

\begin{abstract}
We present preliminary results from high resolution ($\sim 0.05"$)
adaptive optics observations of Cygnus A.
The images show a bi-conic structure strongly suggestive of an obscuring 
torus around a quasar nucleus.
A bright ($K'=18.5$) point source is found near the expected position of the
nucleus.  We interpret this source as the hot inner rim of the torus seen
through the opening of the torus.  Using high angular resolution 
$K$-band spectroscopy, we measure the ratio of molecular to recombination 
hydrogen lines as a function of distance to the center of the putative torus.
These measurements place constraints on the properties of the torus and 
indicate a projected diameter of $\sim 600$ pc.

\end{abstract}


\section{Introduction}
The interplay between dust and radio emission has been the
subject of vigorous research in recent years.   
We now know that radio 
loud sources have a high incidence of dust in their central regions 
(e.g., de Koff et al. 2000) and that
there are several correlations between the properties of the dust and those of
the radio source.  However, this same dust has hampered the study of
the nuclear regions of radio galaxies, 
particularly since most of the high resolution imaging studies 
of the hosts have been carried 
out at optical wavelengths.   

We are conducting a Keck and Lick adaptive optics (AO) imaging and
spectroscopic survey of nearby radio galaxies and other AGN in the near 
infrared that will allow us to pierce the dust
in these objects and study the nuclear regions.  
One of the goals of this survey is to study the obscuring dust tori 
(if present) in these objects.
Here we present preliminary results for one of the 
objects in our sample, the prototype radio galaxy Cyg A.  For details
on the observations, see Canalizo et al. (2003, hereafter Paper I; 2004,
in preparation).

\section[Torus in Cygnus A]
{The Torus in Cygnus A}
Because of its proximity ($z=0.056$) and extreme characteristics, Cyg A 
has played a fundamental role in the study of virtually every aspect of 
powerful radio galaxies.
Different lines of evidence indicate that Cyg A harbors a 
heavily extincted quasar (e.g., Ogle et al. 1997).  According to unification
schemes, this quasar would be hidden by an optically and geometrically thick
dust torus.
Figure~1 shows a Keck NIRC2 AO $K'$ image of the central region of Cyg A
with a resolution of $\sim0.05"$ (or 50 pc for 
$H_{0}=70$ km s$^{-1}$ Mpc$^{-1}$).   The two edge-brightened cones
clearly seen in this image are strongly suggestive of a dust torus casting 
a shadow onto the surrounding gas.  There are two unresolved sources
in the central region: the "primary", near 
(but slightly off-center from) the vertex of the cones, and the "secondary", 
which appears to be the tidally stripped core of a lower mass merging galaxy
(Paper I).

\begin{figure}[ht]
\vskip.2in
\centerline{\includegraphics[width=4.5in]{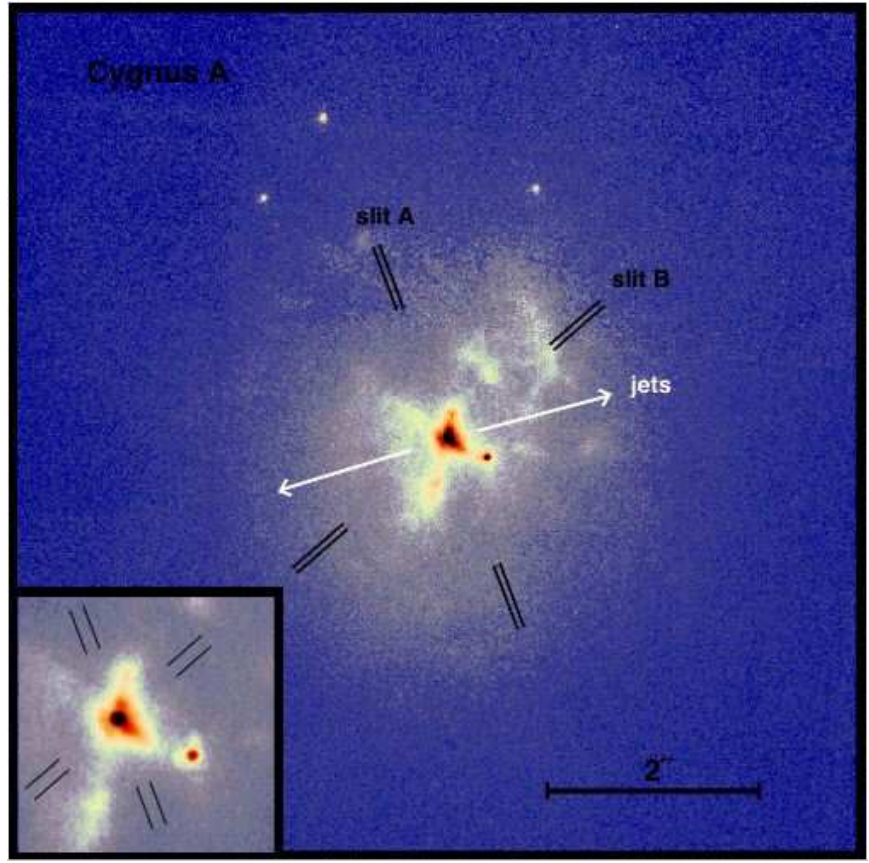}}
\caption{False-color NIRC2 AO $K'$ image of the core of Cyg A. North is up
and East is to the left. White arrows indicate the direction of the jets 
and parallel lines the slit positions.}
\end{figure}

$HST$ NICMOS observations of the primary point source show that it
is highly polarized ($P_k \sim 25\%$) with a polarization angle roughly 
perpendicular to the jets (Tadhunter et al. 2000; 1999).  Overlaying these
data with $HST$ FOC polarization data indicates that the primary
is within 0.1" of the scattering center (M. Kishimoto, personal communication).
While it is tempting to associate this point source with the
quasar nucleus, its flux is considerably higher than that predicted by X-ray
observations.  Ward et al. (1991) predict a continuum flux density at 2.2 
$\mu$m of $3.6\times10^{-15}$ erg cm$^{-2}$ s$^{-1}$ \AA$^{-1}$.  Assuming a 
normal dust-to-gas ratio, Ueno et al. (1994) estimate $A_V = 170$ from the
X-ray absorption column.  Combining both results, we expect a quasar
continuum emission of $9.6\times10^{-21}$ erg cm$^{-2}$ s$^{-1}$ \AA$^{-1}$
at 2.2 $\mu$m.  However, we measure a 2.2 $\mu$m flux of
$1.53\times10^{-18}$ erg cm$^{-2}$ s$^{-1}$ \AA$^{-1}$.   From our NIRC2
$K$-band spectrum (see below; Fig.~2) we estimate that
36\% of the total flux comes from emission lines, so the observed 
continuum flux density is  $9.8\times10^{-19}$ erg cm$^{-2}$ s$^{-1}$ 
\AA$^{-1}$, two orders of magnitude greater than expected.   It is certainly 
possible that the extinction may be overestimated if the mean dust grain size
is larger than assumed (e.g. Maiolino et al. 2001).  However, the $K$-band 
spectrum of this point source (Fig.~2) does not show obvious broad lines, 
indicating that the broad line region is indeed at least partially hidden at 
these wavelengths.

\begin{figure}[ht]
\centerline{\includegraphics[width=4in]{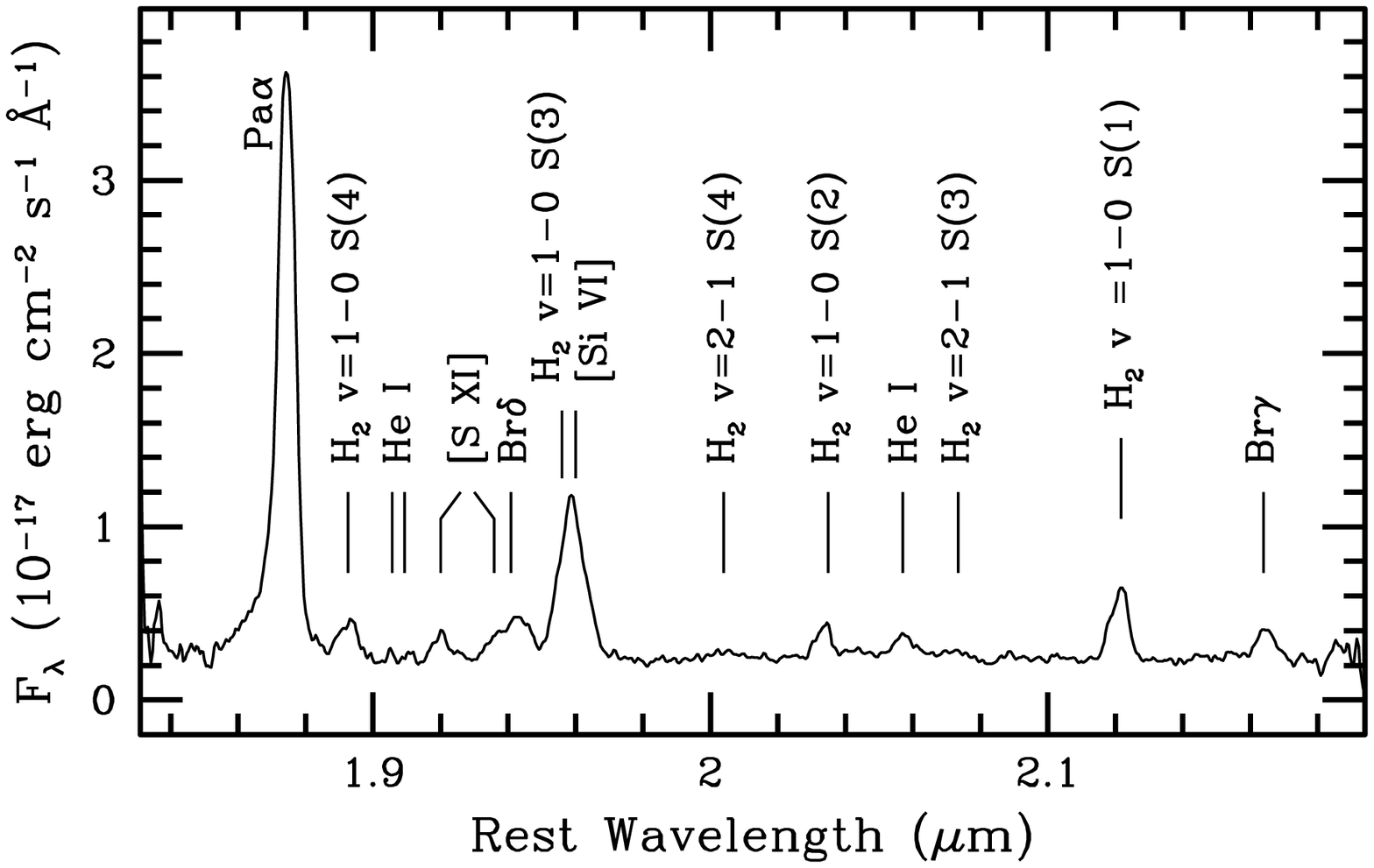}}
\caption{Keck NIRC2 AO spectrum of the 640 central parsecs of Cyg A.
The slit was placed roughly perpendicular to the direction of the jets (Slit A
in Fig.~1)}
\end{figure}

A more likely interpretation for the $K'$ point source is that it is the 
hot inner rim of the dust torus seen through the upper opening of the torus.
The quasar itself is hidden behind the SE half of the torus, 
which is closer to us than its NW counterpart (recall that the 
NW jet points toward us and the SE jet away from us; see e.g., Carilli 
\& Barthel 1996).  Using observations of
the 9.7 $\mu$m silicate dust absorption feature toward the nucleus, Imanishi
\& Ueno (2000) estimate an inner radius for the torus of less than 10 pc. 
Since the resolution of our images is 50 pc, a torus opening of this order
would certainly appear unresolved.

We obtained $K$-band Keck AO spectroscopy at the two slit positions 
indicated in Fig.~1.  Slit A corresponds to the NIRC2 spectrum shown in 
Fig.~2, with total integration of 3600 s.   The slit was placed roughly 
perpendicular to the jets, presumably 
along the torus.   The plate scale in the cross-dispersion direction is 
0.04" pixel$^{-1}$ (or $\sim$ 40 pc for Cyg A).
Slit B corresponds to a NIRSPEC spectrum with a total integration time of
2400 s.  This slit was placed as close to the axis of the jets as possible, 
given the constraints imposed by the AO system (Paper I).
The data were binned in the cross dispersion direction to match the spatial
resolution of the NIRC2 data.

Figure~3 shows, for each slit position, a plot of the flux ratio of the 
H$_2$ $\nu = 1-0$ S(1) line 
to Pa$\alpha$, as a function of the distance from the primary point source.
For Slit A (left panel), the ratio is clearly lower in the region 
centered around the point source.  The change in the ratio is unlikely to 
be due to extinction, since the ratio of Br$\gamma$ to Pa$\alpha$ remains
constant over this region.  A plausible explanation for this trend 
is that the molecular hydrogen is shielded from the continuum source on either
side of the torus, leading to a higher H$_2/$Pa$\alpha$ ratio.   Since
the torus is inclined some 30 to 60 degrees with respect to the line of sight,
the emission we observe in this region actually comes from the region 
{\it above} the torus.  Here the molecular hydrogen is no 
longer shielded and is thus photodissociated, so that the only emission we 
observe comes from the foreground; hence its ratio to Pa$\alpha$ drops.   
From this plot, it may be inferred that the diameter of the torus is between
500 and 600 pc.

\begin{figure}[ht]
\sidebyside
{\centerline{\includegraphics[width=2.7in]{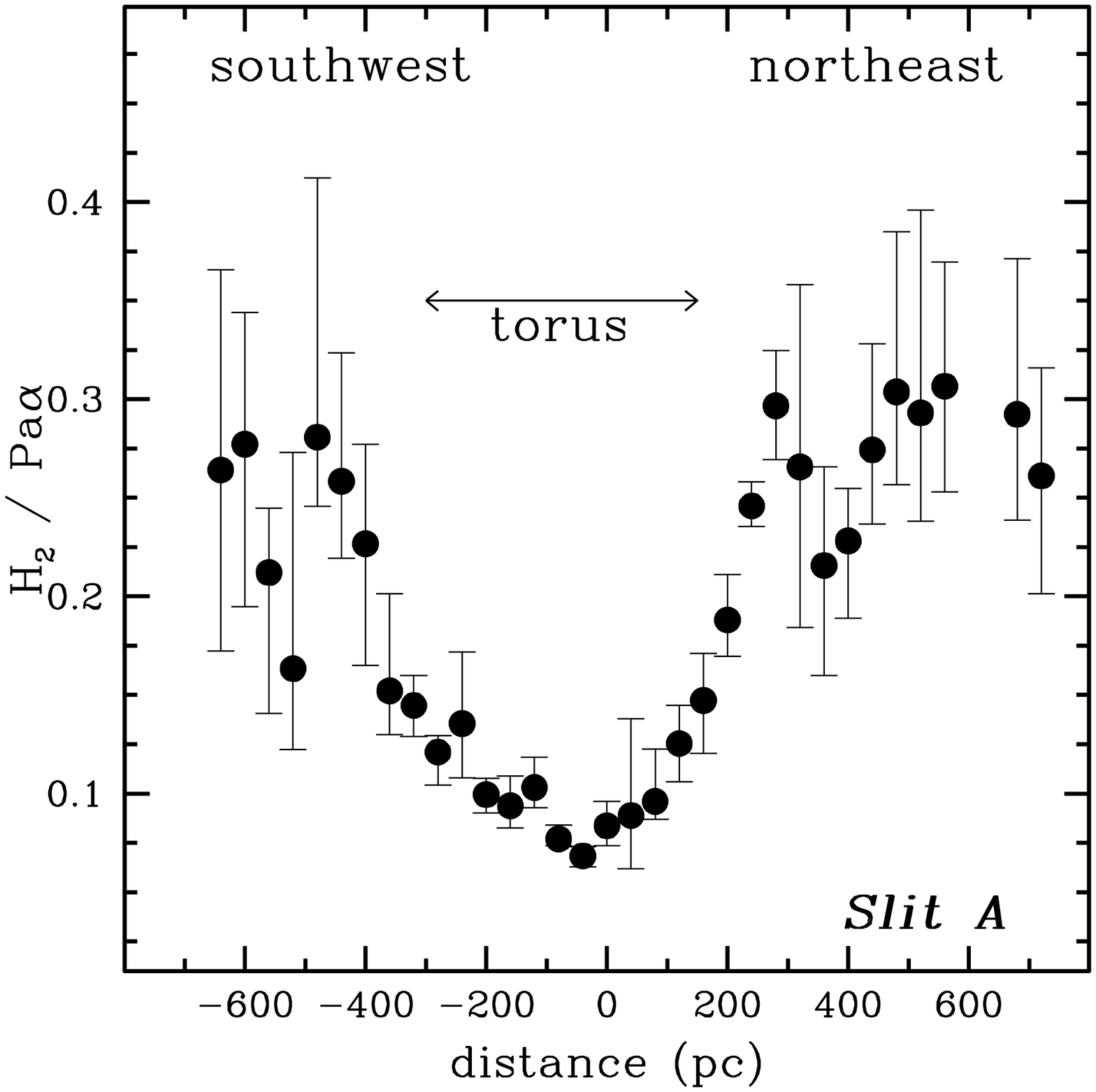}}}
{\centerline{\includegraphics[width=2.7in]{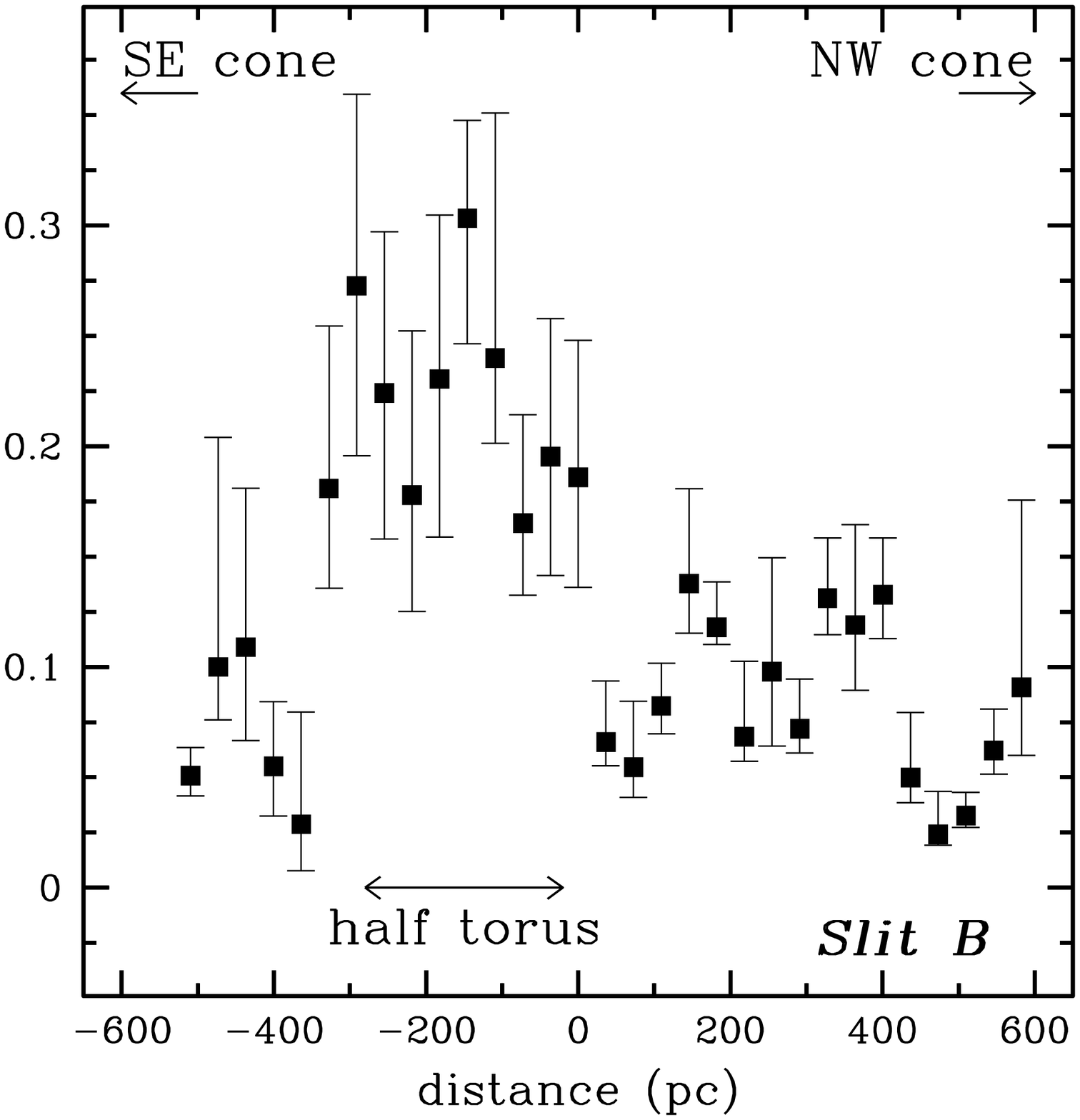}}}
\caption{The ratio of the flux of H$_2$ $\nu = 1-0$ S(1) to Pa$\alpha$ plotted
as a function of distance from the primary point source for Slit A (left) and
Slit B (right).}
\end{figure}

In the case of Slit B (Fig.~3, right panel), we are measuring
H$_2/$Pa$\alpha$ in the direction of the ionization cones, so that H$_2$ is 
mostly photodissociated within the cones.   However, in the
region SE of the opening of the torus (at distance $<0$), the ratio 
increases since this region is shielded by the SE half of the torus.
The ratio decreases again at roughly $-300$ pc, presumably beyond the shadow
of the torus.   The projected size of the torus would then be roughly 
$300\times2=600$ pc.  To obtain an actual diameter of the torus from this value
we must take into consideration the scale height as well 
as the inclination angle, and the angle between the slit and the axis of 
the jets.

Thus, we infer a projected diameter of $\sim$600 pc for the torus in 
Cyg A.  This value is consistent with values inferred using other methods, 
e.g., $< 800$ pc from $\sim0.8"$ resolution optical images 
(Vestergaard \& Barthel 1993) and 
"a few hundred parsecs" from infrared observations of the 9.7 $\mu$m silicate 
dust absorption feature (Imanishi \& Ueno 2000).  Detailed modeling should 
allow us to
obtain better estimates for the size and inclination of the torus.

\begin{acknowledgments}
This work was supported in part under the auspices 
of the U.S.\ Department of Energy, National Nuclear Security 
Administration by the University of California, Lawrence Livermore 
National Laboratory under contract No. W-7405-Eng-48.
\end{acknowledgments}


\begin{chapthebibliography}{1}

\bibitem{canalizo}
Canalizo, G., Max, C. E., Whysong, D., Antonucci, R., Dahm, S. E. 
2003, ApJ, 597, 823

\bibitem{carilli96}
Carilli, C. L., \& Barthel, P. D. 1996, A\&A Rev., 7, 1

\bibitem{deKoff}
de Koff, S. et al. 2000, ApJ, 129, 33

\bibitem{imanishi00}
Imanishi, M., \& Ueno, S. 2000, ApJ, 535, 626


\bibitem{mai01}
Maiolino, R., Marconi, A., Salvati, M., Risaliti, G., Severgnini, P., Oliva, 
E., La Franca, F., Vanzi, L. 2001, A\&A, 365, 28


\bibitem{ogle97}
Ogle, P., et al. 1997, ApJ, 482, L37

\bibitem{tad99} 
Tadhunter, C. N., Packham, C., Axon, D. J., Jackson, N. J., Hough, J. H., 
Robinson, A., Young, S., Sparks, W. 1999, ApJ, 512, 91

\bibitem{tad00} 
Tadhunter, C. N., Sparks, W., Axon, D. J., Bergeron, L., Jackson, N. J., 
Packham, C., Hough, J. H., Robinson, A., Young, S. 2000, MNRAS, 313, 52

\bibitem{ueno94}
Ueno, S., Koyama, K., Nishida, M., Yamauchi, S., Ward, M.J. 1994, ApJL, 431, 1

\bibitem{vester93}
Vestergaard, M., Barthel,P.D. 1993, AJ, 105, 456 

\bibitem{ward}
Ward, M. J., Blanco, P. R., Wilson, A. S., Nishida, M. 1991, ApJ, 382, 115

\end{chapthebibliography}
\end{document}